\documentclass[12pt]{article}
\usepackage{epsf}
\usepackage{cite,,epsfig}
\usepackage{graphicx}

\newcommand{\be}{\begin{equation}}
\newcommand{\ee}{\end{equation}}
\newcommand{\bea}{\begin{eqnarray}}
\newcommand{\eea}{\end{eqnarray}}
\newcommand{\beq}{\begin{equation}}
\newcommand{\eeq}{\end{equation}}
\newcommand{\ba}{\begin{array}}
\newcommand{\ea}{\end{array}}
\newcommand{\beqa}{\begin{eqnarray}}
\newcommand{\eeqa}{\end{eqnarray}}

\newcommand{\cO}{{\cal O}}

\newcommand{\no}{\nonumber}

\newcommand{\lsim}{\stackrel{<}{_\sim}}
\newcommand{\gsim}{\stackrel{>}{_\sim}}

\newcommand{\BR}{{\mathcal B}}

\newcommand{\Mtlr}{\tilde M_{t}^{LR}}

\newcommand{\msq}{M_{\tilde{q}}}
\newcommand{\msl}{M_{\tilde{\ell}}}
\newcommand{\Btaun}{{B_u \to \tau \nu}}

\newcommand{\mg}{M_{\tilde g}}
\newcommand{\msqL}{M_{{\tilde q}_L}}

\newcommand{\msdR}{M_{{\tilde d}_R}}

\newcommand{\msuR}{M_{{\tilde u}_R}}

\def\plb#1#2#3{    {\it Phys. Lett. }{\bf B #1} (#2) #3}

\def\prd#1#2#3{    {\it Phys. Rev. }{\bf D #1} (#2) #3}

\def\prl#1#2#3{    {\it Phys. Rev. Lett. }{\bf #1} (#2) #3}

\def\jhep#1#2#3{   {\it JHEP  }{\bf #1} (#2) #3}

\begin{document}

\thispagestyle{empty}
\begin{flushright}
April 2006 
\end{flushright}
\vskip 1.5 true cm 

\begin{center}
{\Large\bf Hints of large $\tan\beta$ in flavour physics} 
 \\ [25 pt]
{\sc {\sc Gino Isidori${}^{a}$ and Paride Paradisi${}^{b}$ } 
 \\ [25 pt]
{\sl ${}^a$INFN, Laboratori Nazionali di Frascati, Via E. Fermi 40, 
           I-00044 Frascati, Italy} \\ [5 pt] 
{\sl ${}^b$INFN, Sezione di Roma II and Dipartimento di Fisica, \\
           Universit\`a di Roma ``Tor Vergata'', Via della Ricerca Scientifica 1,
	   I-00133 Rome, Italy} \\ 
[25 pt]  }   
{\bf Abstract} \\
\end{center}
\noindent
Motivated by the first evidence of the $\Btaun$
transition reported by Belle \cite{Btn_Belle} and by 
the precise $\Delta M_{B_s}$ measurement by CDF \cite{DMs_CDF}, 
we analyse these and other low-energy observables in the 
framework of the MSSM at large $\tan\beta$. We show that for
heavy squarks and $A$ terms ($\msq,~A_U~\gsim 1$~TeV)
such scenario has several interesting virtues.
It naturally describes: i)~a suppression of 
$\BR(\Btaun)$ of (10-40)\%, ii) a sizable enhancement of $(g-2)_\mu$,  
iii) a heavy SM-like Higgs ($m_{h^0} \sim 120$ GeV),
iv) small non-standard effects in 
$\Delta M_{B_s}$ and $\BR(B\to X_s \gamma)$
(in agreement with present observations).
The possibilities to find more convincing 
evidences of such scenario, with improved data on
$\BR(\Btaun)$, $\BR(B_{s,d}\to \ell^+\ell^-)$
and other low-energy observables,
are briefly discussed. 

\vskip 1.5 cm

\section{Introduction}
In many extensions of the SM, including the so-called Minimal 
Supersymmetric extension of the SM (MSSM), the Higgs sector 
consists of two  $SU(2)_L$ scalar doublets, coupled 
separately to up- and down-type quarks. 
A key parameter of all these models is 
$\tan\beta=v_u/v_d$, the ratio 
of the two Higgs vacuum expectation values.
This parameter controls the overall normalization 
of the Yukawa couplings. The regime of large $\tan\beta$ 
[$\tan\beta = \cO(m_t/m_b)$] has an intrinsic theoretical 
interest since it allows the unification of 
top and bottom Yukawa couplings, as predicted 
in well-motivated grand-unified models \cite{GUT}.

The large $\tan\beta$ regime of both supersymmetric and 
non-supersymmetric models has a few interesting 
signatures in $B$ physics. One of the most clear ones
is the suppression of $\BR(\Btaun)$
with respect to its SM expectation \cite{Hou}.
Potentially sizable effects are expected also in  
$\BR(B\to X_s \gamma)$, $\Delta M_{B_s}$ and 
$\BR(B_{s,d}\to \ell^+\ell^-)$. 
Motivated by the recent experimental results on both 
$\BR(\Btaun)$ \cite{Btn_Belle} and 
$\Delta M_{B_s}$ \cite{DMs_CDF}, 
we present here a correlated analysis of all these observables 
within the large $\tan\beta$ limit of the MSSM. 
 
Because of the effective non-holomorphic terms which break the Peccei-Quinn 
symmetry of the tree-level Yukawa interaction \cite{HRS,BRP},
the phenomenology of the  MSSM in the large $\tan\beta$ regime is richer
than in non-supersymmetric models. 
We pay particular attention to re-summation effects beyond the one-loop 
level, both in charged- and in neutral-current interactions, 
which  play a key role in the correlations among these $B$-physics 
observables \cite{Carena,bsgamma,Babu,IR,Buras,vives}.

The generic MSSM contains in principle several free 
parameters in addition to $\tan\beta$. Given the 
absence of significant non-standard effects 
both in the electroweak and in the flavour sector, 
we limit ourselves to a Minimal Flavour Violating (MFV) scenario 
\cite{MFV_randall,MFV} with squark masses 
in the TeV range. In addition, we take 
into account the important information on the model 
derived by two flavour-conserving observables:
the anomalous magnetic moment of the muon 
and the lower limit on the lightest Higgs boson mass.

The present central values of the measurements of 
$\BR(\Btaun)$ and $(g-2)_\mu$  are substantially different 
from the corresponding SM expectations. Although both 
these effects are not statistically significant yet, 
we find that these central values can naturally be 
accommodated within this scenario (for a wide range 
of $\mu$, $\tan\beta$ and the charged Higgs mass).
More interestingly, if the trilinear term $A_U$ 
is sufficiently large, this scenario can also explain why the 
lightest Higgs boson has not been observed yet.
Finally, the parameter space which leads to these 
interesting effects can also naturally explain why 
$\BR(B\to X_s \gamma)$ and $\Delta M_{B_s}$ are
in good agreement with the SM expectations. 
We are therefore led to the conclusion that,
within the supersymmetric extensions of the SM, 
the scenario with large $\tan\beta$ and heavy 
soft-breaking terms in the squark sector 
is one of the most interesting and likely possibilities.

The plan of the paper is the following: in Section~\ref{sect:Bphys}
we recall the basic formulae to analyse large-$\tan\beta$
effects in  $\BR(\Btaun)$, $\BR(B_{s,d}\to \ell^+\ell^-)$,
$\Delta M_{B_s}$, and $\BR(B\to X_s \gamma)$. We pay particular attention 
to the $\BR(\Btaun)$ case, analysing the resummation of 
large $\tan\beta$ effects beyond the lowest order and the 
strategy to decrease the theoretical uncertainty with the 
help of $\Delta M_{B_d}$. In Section~\ref{sect:mg_gm2}
we discuss the implications on the MSSM parameter space 
derived by $m_{h^0}$ and $(g-2)_\mu$. The correlated 
analysis of all the observables is presented 
in  Section~\ref{sect:discuss}, together with a discussion 
about future tests of the model by means of 
other $P\to \ell\nu$ decays. The results 
are summarized in the Conclusions. 

\section{$B$-physics observables}
\label{sect:Bphys}

\subsection{ $\Btaun$}
The SM expectation for the $\Btaun$
branching fraction is 
\begin{equation}
 \label{eq:BR_B_taunu}
{\cal B}(\Btaun)^{\rm SM} = 
\frac{G_{F}^{2}m_{B}m_{\tau}^{2}}{8\pi}\left(1-\frac{m_{\tau}^{2}}
{m_{B}^{2}}\right)^{2}f_{B}^{2}|V_{ub}|^{2}\tau_{B}~.
\end{equation}
Using  $|V_{ub}| = (4.39 \pm 0.33) \times 10^{-3}$ from inclusive
$b\to u$
semileptonic decays \cite{HFAG}, 
$\tau_{B} = 1.643\pm 0.010$~ps\cite{PDG},
and the recent unquenched lattice result $f_B = 0.216\pm 0.022$ GeV \cite{Lattice_fb}, 
this implies ${\cal B}(\Btaun)^{\rm SM}=(1.59\pm 0.40) \times 10^{-4}$.
This prediction should be compared with Belle's recent result~\cite{Btn_Belle}:
\begin{equation}
\label{bellebtnu}
{\cal B}(B^- \to \tau^-  \bar \nu)^{\rm exp} = 
(1.06^{+0.34}_{-0.28}(\mbox{stat})^{+0.18}_{-0.16}(\mbox{syst}))\times 10^{-4}~.
\end{equation}

Within two-Higgs doublet models, the charged-Higgs exchange amplitude 
induces an additional tree-level contribution to semileptonic decays.
Being proportional to the Yu\-ka\-wa couplings of quarks and leptons,
this additional contribution is usually negligible.
However, in $B\to\ell \nu$ decays the $H^{\pm}$ exchange 
can compete with the $W^{\pm}$ exchange thanks to the 
helicity suppression of the latter.  
Interestingly, in models where the two Higgs doublets are
coupled separately to up- and down-type quarks,
the interference between $W^{\pm}$ and $H^{\pm}$ 
amplitudes is necessarily {\em destructive} \cite{Hou}.

Taking into account the resummation of the leading $\tan\beta$
corrections to all orders, the charged-Higgs contributions to 
the $\Btaun$  amplitude within a MFV supersymmetric framework 
lead to the following ratio:
\beq
R_{B\tau\nu} = \frac{\BR^{\rm SUSY}(\Btaun)}{\BR^{\rm SM}(\Btaun)}
=\left[1-\left(\frac{m^{2}_B}{m^{2}_{H^\pm}}\right)
\frac{\tan^2\beta}{(1+\epsilon_0\tan\beta)}
\right]^2~,
\label{eq:Btn}
\eeq
where $\epsilon_0$ denotes the effective coupling which 
parametrizes the non-holomorphic correction to the down-type Yukawa coupling 
induced by gluino exchange (see Section~\ref{sect:B_others}). We stress that 
the result in Eq.~(\ref{eq:Btn})
takes into account all the leading $\tan\beta$
corrections both in the redefinition of the 
bottom-quark Yukawa coupling and in the 
redefinition of the CKM matrix.\footnote{The result in Eq.~(\ref{eq:Btn})
can easily be obtained by means of the charged-Higgs effective 
Lagrangian in Eq.~(52) of Ref.~\cite{MFV}, which systematically
takes into account the redefinition of 
Yukawa couplings and CKM matrix elements.
The explicit application to $\Btaun$ has been presented first in Ref.~\cite{btnu}.}

For a natural choice of the parameters 
($30\lsim \tan\beta\lsim 50$, $0.5 \lsim M_{H^\pm}/\rm{TeV} \lsim 1$, $\epsilon_0\sim 10^{-2}$)
Eq.~(\ref{eq:Btn}) implies a (5-30)\% suppression 
with respect to the SM. This would perfectly fit with 
the experimental result in (\ref{bellebtnu}), which implies
\begin{equation}
R_{B\tau\nu}^{\rm exp} =
\frac{\BR^{\rm exp}(\Btaun)}{\BR^{\rm SM}(\Btaun)} 
~=~ 0.67^{+0.30}_{-0.27} 
~=~ 0.67^{+0.24}_{-0.21~{\rm exp}} \pm 0.14_{|f_{B}|} \pm 0.10_{|V_{ub}| }~.
\label{eq:Rtn_exp}
\end{equation}

Apart from the experimental error, one of the difficulties 
in obtaining a clear evidence of a possible deviation of $R_{B\tau\nu}$ from unity
is the large parametric uncertainty induced by $|f_{B}|$ and $|V_{ub}|$.
As suggested by Ikado~\cite{Belle_Btn/DM}, an interesting way to partially 
circumvent this problem 
is obtained by normalizing $\BR(\Btaun)$
to the $B_d$--$\bar B_d$ mass difference ($\Delta M_{B_d}$). 
Neglecting 
isospin-breaking differences in masses and decay constants
between $B_d$ and $B_u$ mesons, we can write  
\beqa
\left. \frac{\BR(\Btaun)}{\tau_{B_u} \Delta M_{B_d} } \right|^{\rm SM} &=& 
\frac{3\pi}{4 \eta_B(m_b) S_0(m_t^2/M_W^2) B_{B_d}(m_b) } \frac{m^2_\tau}{M_W^2}
\left(1-\frac{m^{2}_\tau}{m^{2}_{B}}\right)^2 \left|\frac{V_{ub}}{V_{td}}\right|^2~, \\
&=& 1.24 \times 10^{-4} \left( \frac{|V_{ub}/V_{td}|}{0.473}\right)^2 
\left(\frac{0.836}{B_{B_d}(m_b)}\right)~. 
\label{eq:Btn_DMB}
\eeqa
Following standard notations,  
we have denoted by $S_0$, $\eta_B$ and $B_{B_d}$
the Wilson coefficient, the QCD correction factor and the bag parameter
of the $\Delta B=2$ operator within the SM (see e.g.~Ref.\cite{Buras}).
Using the unquenched lattice result $B_{B_d}(m_b) = 0.836 \pm 0.068$ \cite{Aoki} 
and $|V_{ub}/V_{td}|=0.473 \pm 0.024 $ from the UTfit Collaboration \cite{Bona},
we then obtain 
\bea
\left(R^\prime_{B\tau\nu}\right)^{\rm exp} &=& 
\frac{\BR^{\rm exp}(\Btaun)/\Delta M_{B_d}^{\rm exp}}{\BR^{\rm SM}(\Btaun)/\Delta M_{B_d}^{\rm SM}} 
\\
&=& 1.03^{+0.39}_{-0.33} 
~=~ 1.03^{+0.37}_{-0.31~{\rm exp}} \pm 0.08_{|B_{B_d}|} \pm 0.10_{|V_{ub}/V_{td}| }~.
\label{eq:Rtn_prime}
\eea
The following comments follow from the comparison of 
Eqs.~(\ref{eq:Rtn_exp}) and (\ref{eq:Rtn_prime}):
\begin{itemize}
\item  
The two results are compatible and with similar overall relative errors.
However, the parametric/theoretical component is smaller in Eq.~(\ref{eq:Rtn_prime}). 
The latter could therefore become a more stringent test of the SM in the near future, 
with higher statistics on the $\Btaun$ channel. 
\item In generic extensions of the SM,  $R_{B\tau\nu}$ and 
$R^\prime_{B\tau\nu}$ are not necessarily the same. However, they should
coincide if the non-SM contribution to $\Delta M_{B_d}$ is negligible, which is
an excellent approximation in the class of models we are considering here.
\item For consistency, the $|V_{ub}/V_{td}|$ combination entering in 
 Eq.~(\ref{eq:Rtn_prime})
should be determined without using the information on $\Delta M_{B_d}$
and $\Btaun$ (condition which is already almost fulfilled). 
In the near future one could determine this ratio with 
negligible hadronic uncertainties using the 
relation $|V_{ub}/V_{td}|=|\sin\beta_{_{\rm CKM}}/\sin\gamma_{_{\rm CKM}}|$.
\end{itemize}

\subsection{$\BR(B_{s,d} \to \ell^+ \ell^-)$, $\BR(B\to X_s \gamma)$, 
and $B_{s}$--${\bar B}_{s}$ mixing}
\label{sect:B_others}

The important role of these observables in the MSSM with MFV and
large $\tan\beta$ has been widely discussed in the literature
\cite{bsgamma,Babu,IR,Buras} (see also \cite{others_mm,Nierste,Carena_new,Buras_new}).
We recall here only a few ingredients which are necessary to 
analyse their correlations with $\BR(\Btaun)$. 

We are interested in a scenario with heavy squarks, 
where the $SU(2)_L$-breaking corrections of $\cO(M_W/\msq)$
can be treated as a small perturbation. In this limit, 
the $\tan\beta$-enhanced corrections to the down-type Yukawa 
couplings are parameterized by the following effective couplings \cite{HRS}
\beq
\epsilon_0 = - \frac{ 2 \alpha_{\rm s}\mu }{3\pi\mg } 
 H_2 \left( \frac{\msqL^2}{\mg^2}, \frac{\msdR^2}{\mg^2} \right)~, \qquad 
\epsilon_Y = - \frac{A_U }{16 \pi^2 \mu }  
 H_2 \left( \frac{\msqL^2}{\mu^2}, \frac{\msuR^2}{\mu^2} \right)~, \qquad 
\eeq
where
\be
H_2(x,y) = \frac{x\ln\, x}{(1-x)(x-y)} + \frac{y\ln\, y}{(1-y) (y-x)}
\ee
and, as usual, $\mu$ denotes the supersymmetric Higgs mass term
and $A_U$ the three-linear soft-breaking term.

In $\BR(B_{s,d} \to \ell^+ \ell^-)$ and $B_{s}$--${\bar B}_{s}$ the
only relevant contributions in the limit of heavy squarks 
are the effective tree-level Higgs-mediated neutral currents.
In the $\BR(B_{s,d} \to \ell^+ \ell^-)$ case this leads to \cite{Babu,IR,MFV}:
\bea
R_{B\ell\ell} &=&
\frac{ \BR^{\rm SUSY}(B_q \to \ell^+ \ell^-) }{ \BR^{\rm SM}(B_q \to \ell^+ \ell^-) } =
      (1+\delta_S)^2 +
       \left(1-\frac{4 m^2_\ell}{M^2_{B_q }}\right) \delta_S^2~, 
\label{eq:Bmm} \\
\delta_S &=& \frac{\pi \sin^2\theta_w M^2_{B_q }  }{ \alpha_{\rm em} M_A^2 C_{10A}(m_t^2/M_W^2) }
 \frac{\epsilon_Y \lambda_t^2 \tan^3\beta }{[1+(\epsilon_0  +  \epsilon_Y \lambda_t^2)\tan\beta]
[1 + \epsilon_0  \tan\beta]}~, \qquad 
\eea
where $\lambda_t$ is the top-quark Yukawa coupling, $C_{10A}(m_t^2/M_W^2)$ is the SM Wilson
coefficient  ($\lambda_t\approx 1$, $C_{10A}\approx 1$) and 
$M_A$ is the mass of the physical pseudoscalar Higgs 
(at the tree level $M^2_A=M^2_{H^\pm}-M_W^2$). 
As discussed in \cite{Babu,IR,Buras,others_mm}, 
for $\tan\beta \sim 50$ and $M_A \sim 0.5$~TeV
the neutral-Higgs contribution to $\BR(B_{s,d} \to \ell^+ \ell^-)$
can easily lead to an $\cO(100)$ enhancement over the SM expectation.
This possibility is already excluded by experiments: the CDF  
bound $\BR(B_s \to \mu^+\mu^-)< 8.0 \times 10^{-8}$ \cite{Bmm}
implies 
\be
R_{B\ell\ell} < 23 \qquad  [90\%~{\rm C.L.}]~.
\label{eq:Bll_lim}
\ee
As we will discuss in Section~\ref{sect:discuss}, 
this limit poses severe constraints on the MSSM parameter space;
however, it does not prevent a sizable charged-Higgs 
contribution to $\BR(\Btaun)$. This can easily be understood 
by noting that the effect in Eq.~(\ref{eq:Bmm}) vanishes 
for $A_U \to 0$ and has a stronger dependence on $\tan\beta$
than $R_{B\tau\nu}$.

The neutral-Higgs contribution to $\Delta M_{B_{s}}$ leads to \cite{Buras,IR,MFV}:
\bea
R_{\Delta M_s} = 
\frac{(\Delta M_{B_s})^{\rm SUSY}}{(\Delta M_{B_s})^{\rm SM}} 
&=& 1 - m_b(\mu^2_b)m_s(\mu^2_b)
\frac{ 64 \pi \sin^2\theta_w }{\alpha_{\rm em} M^2_A S_0(m_t^2/M_W^2) } \no\\
&& \times \frac{(\epsilon_Y \lambda_t^2 \tan^2\beta)^2 }{[1+(\epsilon_0  +  \epsilon_Y \lambda_t^2)\tan\beta]^2
[1 + \epsilon_0  \tan\beta]^2}~. \qquad 
\label{eq:DMSth}
\eea
where $m_{b,s}(\mu^2_b)$ denotes the bottom- and 
strange-quark masses renormalized  at a scale $\mu_b \approx m_b$.
\footnote{~For simplicity we have set to one the ratio of bag parameters 
between SM and scalar-current operators.} 
The parametric dependence on $A_U$ and $\tan\beta$ in (\ref{eq:DMSth}) 
and (\ref{eq:Bmm}) is quite similar, but the  $m_s(\mu^2_b)$ factor implies a
much smaller non-standard effect in $\Delta M_{B_{s}}$
(typically of a few \%).  Note that,
similarly to the $\BR(\Btaun)$ case, also in $\Delta M_{B_{s}}$ 
one expects a {\em negative} correction with respect 
to the SM. As we will show, thanks to the high experimental 
precision achieved by CDF \cite{DMs_CDF}, at present 
$\Delta M_{B_{s}}$ is comparable
with $\BR(B_{s} \to \mu^+ \mu^-)$ in setting bounds 
in the MSSM parameter space.  
According to the SM expectation 
$(\Delta M_{B_{s}})^{\rm SM} = 21.5 \pm 2.6~{\rm ps}^{-1}$ of the UTfit Collaboration \cite{Bona}, 
the CDF result $(\Delta M_{B_{s}})^{\rm exp}= 17.35 \pm 0.25~{\rm ps}^{-1}$ implies
\be
R_{\Delta M_s}^{\rm exp} =
\frac{(\Delta M_{B_{s}})^{\rm exp}}{(\Delta M_{B_{s}})^{\rm SM}} 
~=~ 0.80 \pm 0.12~. 
\label{eq:DMS_exp}
\ee

The last $B$-physics observable we will consider is $\BR(B\to X_s \gamma)$. 
This observable is particularly sensitive to possible non-standard 
contributions. However, contrary to $\Btaun$, $B_{s,d}\to \ell^+\ell^-$, 
and $\Delta M_{B_s}$,  there is no effective 
tree-level contribution by charged- or neutral-Higgs exchange
in $B\to X_s \gamma$.
Non-standard contributions from the Higgs sector appear only 
at the one-loop level and are not necessarily dominant with respect 
to the chargino-squark contributions, even for squark masses of $\cO(1~{\rm TeV})$.
In the numerical analysis presented in Section~\ref{sect:discuss}
we have implemented the improved chargino-squark amplitude
computed in Ref.~\cite{bsgamma} and the charged- and neutral-Higgs 
exchange amplitudes of  Ref.~\cite{MFV}. A key point to note
is that for $A_U<0$ charged-Higgs and  chargino amplitudes 
tend to cancel themselves. 
This leads to a particularly favorable situation,
given the small room for new physics in this observable.
According to the SM estimate 
$\BR(B\to X_s \gamma)^{\rm SM} = (3.70\pm 0.30)\times 10^{-4}$ 
by Gambino and Misiak \cite{MisiakG},
and the world average $\BR(B\to X_s \gamma)^{\rm exp} 
= (3.52\pm 0.30)\times 10^{-4}$ \cite{HFAG},
we set 
\be
0.76 < R_{B X_s \gamma} = 
\frac{ \BR(B\to X_s \gamma)^{\rm SUSY}}{ \BR(B\to X_s \gamma)^{\rm SM}} < 1.15 
\qquad  [90\%~{\rm C.L.}]~.
\label{eq:bsg_lim}
\ee

\section{Flavour-conserving observables: $m_{h^0}$ and $(g-2)_\mu$}
\label{sect:mg_gm2}

In the previous section we have analysed the impact of a 
MFV supersymmetric scenario with large $\tan\beta$ on 
various $B$-physics observables. As we have seen, 
large $\tan\beta$ values can lead to huge enhancements 
of $\BR(B_{s,d}\to \ell^+\ell^-)$, or a visible 
depression of $\Delta M_{B_s}$, which are
already strongly constrained from data. However, 
the correlation between the large $\tan\beta$ effects 
in these observables crucially depends on magnitude 
(and sign) of the trilinear term  $A_U$.
The question we address in this Section is which 
kind of additional information we can extract on $A_U$
and $\tan\beta$ from two key flavour-conserving 
observables, the lightest Higgs boson mass and the 
anomalous magnetic moment of the muon.

\subsection{The lightest Higgs boson mass}
\label{sect:mhh}

One of the most suggestive prediction of the MSSM 
is the existence of a relatively light neutral Higgs boson ($h^0$)
with $m_{h^0}\lsim 135~\rm{GeV}$ \cite{mh_SUSY}.  
One of the most serious consistency 
problem of the model is why this particle has not 
been observed yet \cite{mh_exp}. 

Even if the $h^0$ mass depends on the whole set of the MSSM
parameters (after the inclusions of loop corrections), 
it is well known that $m_{h^0}$ mainly depends on the
left-right mixing term in the stop mass matrix 
$m_t\Mtlr = m_t(A_U-\mu/\tan\beta)\simeq m_t A_U$ 
(for large $\tan\beta$ values), on the average stop mass 
(which we identify with the average squark mass, $\msq$), 
and on $\tan\beta$.

\begin{figure}[t]
\begin{center}
\hspace{-0.3 cm}
\includegraphics[scale=0.38]{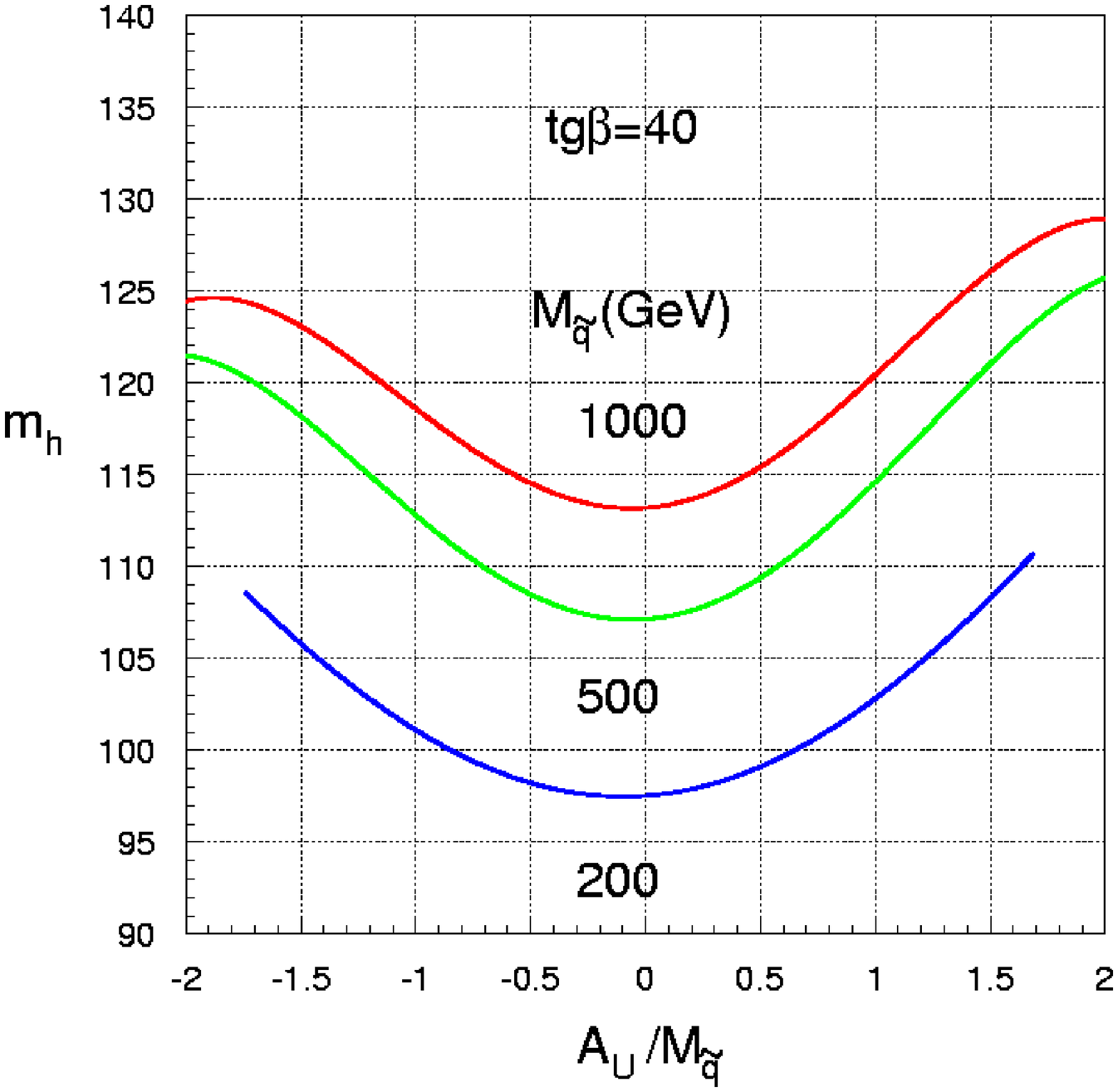}
\hspace{-0.3 cm}
\includegraphics[scale=0.38]{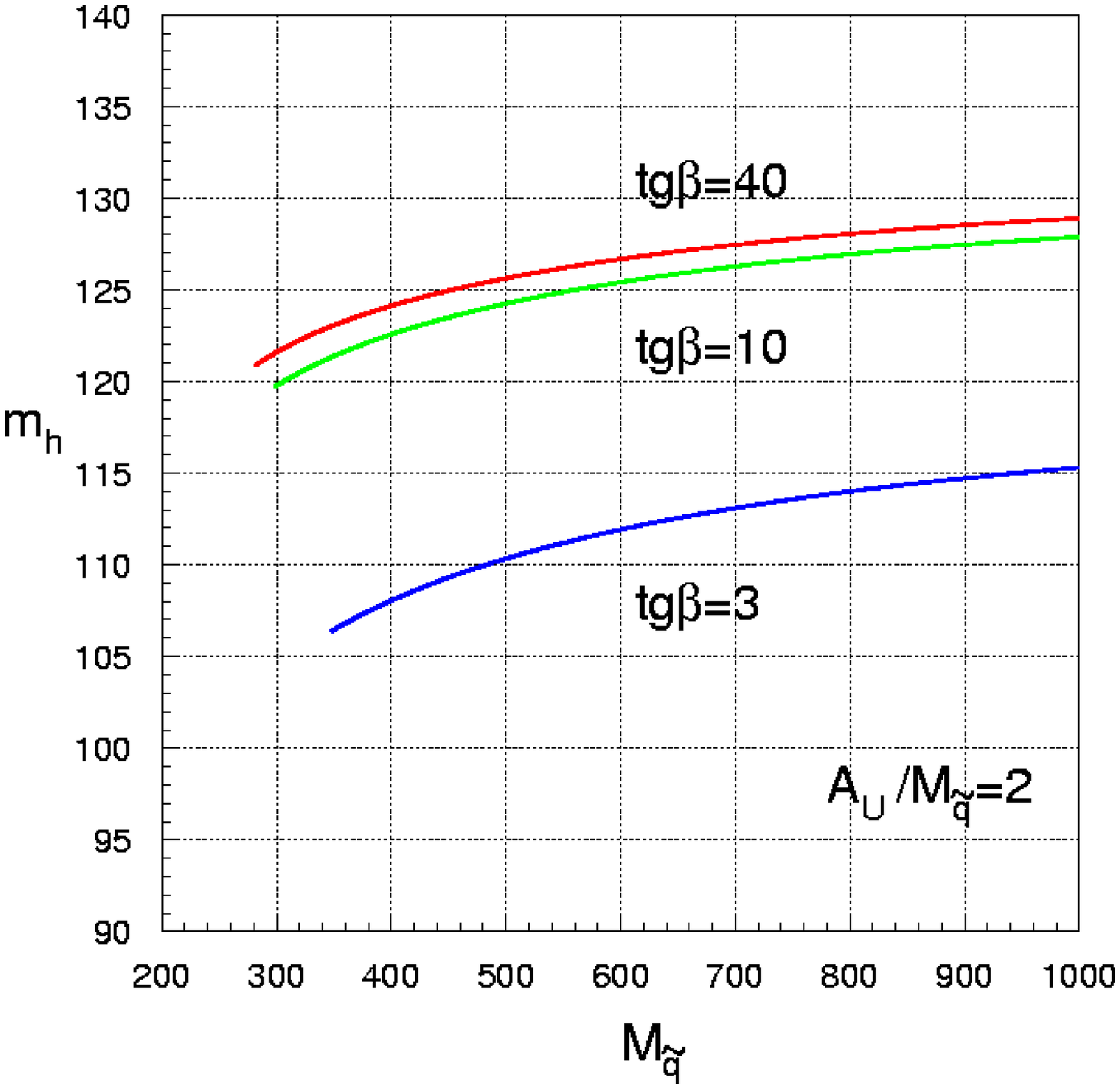}
\hspace{-0.3 cm}\\
\vspace{0.5 cm}
\caption{\label{fig:Higgs} Dependence of the lightest Higgs boson mass
on the average squark mass ($\msq$), $\tan\beta$, and $A_U$.}
\end{center}
\end{figure}

In Fig.~\ref{fig:Higgs} (left) we show $m_{h^0}$ as a function of $A_U/\msq$ 
for $\msq = 200, 500, 1000$~GeV and $M_A = 500~\rm{GeV}$.
A maximum for $m_{h^0}$ is reached for about $A_U/\msq\approx\pm 2$, 
which is usually denoted as the ``maximal mixing'' case.
A minimum is reached around $A_U/\msq\approx 0$, 
which we refer to as the ``no mixing'' case.
As can be seen, even with the most favorable choice of 
$\msq$ and $\tan\beta$, relatively large values 
for $m_{h^0}\geq 120 \rm{GeV}$ are possible only 
if $A_U\gsim \msq$.

The maximal mixing case certainly favours large effects 
in $\BR(B_{s,d}\to \ell^+\ell^-)$ (we recall that 
$\epsilon_{Y} \sim \mu A_U/\msq^2$).
However, it is not possible to establish a well defined 
correlation between $\BR(B_{s,d}\to \ell^+\ell^-)$ and $m_{h^0}$
since  $\BR(B_{s,d}\to \ell^+\ell^-)^{\rm SUSY} \sim \tan^{6}\beta$
while $m_{h^0}$ is rather insensitive 
to $\tan\beta$ for $\tan\beta\geq 10$ (as clearly shown in the right 
plot of Fig.~\ref{fig:Higgs}). Moreover, supersymmetric effects 
in $\BR(B_{s,d}\to \ell^+\ell^-)$ decouple as $M^{4}_{A}$ 
and are proportional to $\mu$, while the $m_{h^0}$ dependence 
on $\mu$ and $M_A$ is quite mild.
In Fig.~\ref{fig:Higgs} (right) we show $m_{h^0}$ as a function 
of $\msq$ in the maximal mixing case for different $\tan\beta$ 
values. 

The two plots in Fig.~\ref{fig:Higgs}
have been obtained including the full one-loop 
corrections and the leading two-loop ones, 
cross-checking the results with those of FeynHiggs \cite{FeynHiggs}.
In summary, we find that in the absence of fine-tuned solutions,
the present experimental lower bounds on $m_{h^0}$ 
\cite{mh_exp} provide a strong support in favour of  
heavy squarks, $A_U\gsim \msq$, and $\tan\beta$ well above unity.
A scenario which enhances the correlations between 
$\BR(\Btaun)$, $\BR(B_{s,d}\to \ell^+\ell^-)$,
$\Delta M_{B_s}$, and $\BR(B\to X_s \gamma)$ at 
large $\tan\beta$. We stress that this 
scenario is quite different from a 
non-supersymmetric two-Higgs doublet model,
even in the limit where $\msq^2 \gg M^2_{H^{\pm}}$.

\subsection{$(g-2)_{\mu}$}

The possibility that the anomalous magnetic 
moment of the muon [$a_\mu = (g-2)_{\mu}/2$],
which has been measured very precisely in the last few years
\cite{g_2_exp}, provides a first hint of physics beyond the SM 
has been widely discussed in the recent literature. 
Despite substantial progress both on the experimental 
and on the theoretical sides, the situation is not completely 
clear yet (see Ref.~\cite{g_2_th} for an updated discussion). 
Most recent analyses converge towards a $2\sigma$ 
discrepancy in the $10^{-9}$ range \cite{g_2_th}:
\be
 \Delta a_{\mu} =  a_{\mu}^{\rm exp} - a_{\mu}^{\rm SM}  
\approx (2 \pm 1) \times 10^{-9}~.
\label{eq:amu_exp}
\ee
If confirmed and interpreted within the MSSM, this result 
would unambiguously signal a large value of $\tan\beta$ 
(see Ref.~\cite{g_2_SUSY1,g_2_mw} and references therein).

The main SUSY contribution to $a^{\rm MSSM}_\mu$ is usually 
provided by the loop exchange of charginos 
and sneutrinos [$(a^{\rm MSSM}_{\mu})_{\chi}\sim 
\alpha_{2} M_{2}\tan\beta/\mu M^{2}_{\tilde{\nu}}$]. 
But if 
$\mu$ is very large --the scenario where ${\mathcal B}\,(B_s \to \mu^+ \mu^-)$
gets its maximum value-- then $a^{\rm MSSM}_\mu$ turns out 
to be dominated by the neutralino (Bino type) amplitude
[$(a^{\rm MSSM}_\mu)_{B}\sim \alpha_1 M_1\mu\tan\beta/ M^{4}_{\tilde{\ell}}$].
A useful tool to illustrate basic features of the supersymmetric 
contribution to $a_\mu$ is the expression
\beq
\frac{a^{\rm MSSM}_\mu}{ 1 \times 10^{-9}}  \approx
2.5  \left(\frac{\tan\beta }{50} \right) 
\left( \frac{500~\rm GeV}{M_{\tilde \chi}} \right)^2
\left( \frac{M_{\tilde \nu}}{M_{\tilde \chi}} \right)~,
\label{eq:g_2}
\eeq
which provides a good approximation to the full
one-loop result \cite{g_2_mw} in the limit of almost degenerate higgsinos  and 
electroweak gauginos ($M_1 \sim M_2 \sim \mu \gg M_W$),
and allowing a moderate splitting between slepton and chargino masses.
>From this expression it is clear that 
values of $a^{\rm MSSM}_\mu$ in the $10^{-9}$ range, 
as suggested by Eq.~(\ref{eq:amu_exp}), require large $\tan\beta$ values. 

As pointed out in \cite{Nierste}, there is a very 
stringent  correlation between $a^{\rm MSSM}_\mu$ and 
$\BR(B_s \to \mu^+ \mu^-)^{\rm MSSM}$ in specific frameworks, 
such as the constrained minimal supergravity scenario. 
However, this correlation is much weaker in a more general 
context, such as the one we are considering here. 
Given the mild $\tan\beta$ dependence of $a_\mu^{\rm MSSM}$
compared to  $\BR(B_s \to \mu^+ \mu^-)^{\rm MSSM}$,
it is easy to generate a sizable contribution to 
$a_\mu$ while keeping  $\BR(B_s \to \mu^+ \mu^-)$
well below its actual experimental resolution.
As we will illustrate in the next section,
there is a stronger model-independent correlation between  
$a_\mu^{\rm MSSM}$ and $\BR(\Btaun)^{\rm MSSM}$.

\begin{figure}[p]
\begin{center}
\hspace{-0.3 cm}
\includegraphics[scale=0.46,angle=-90]{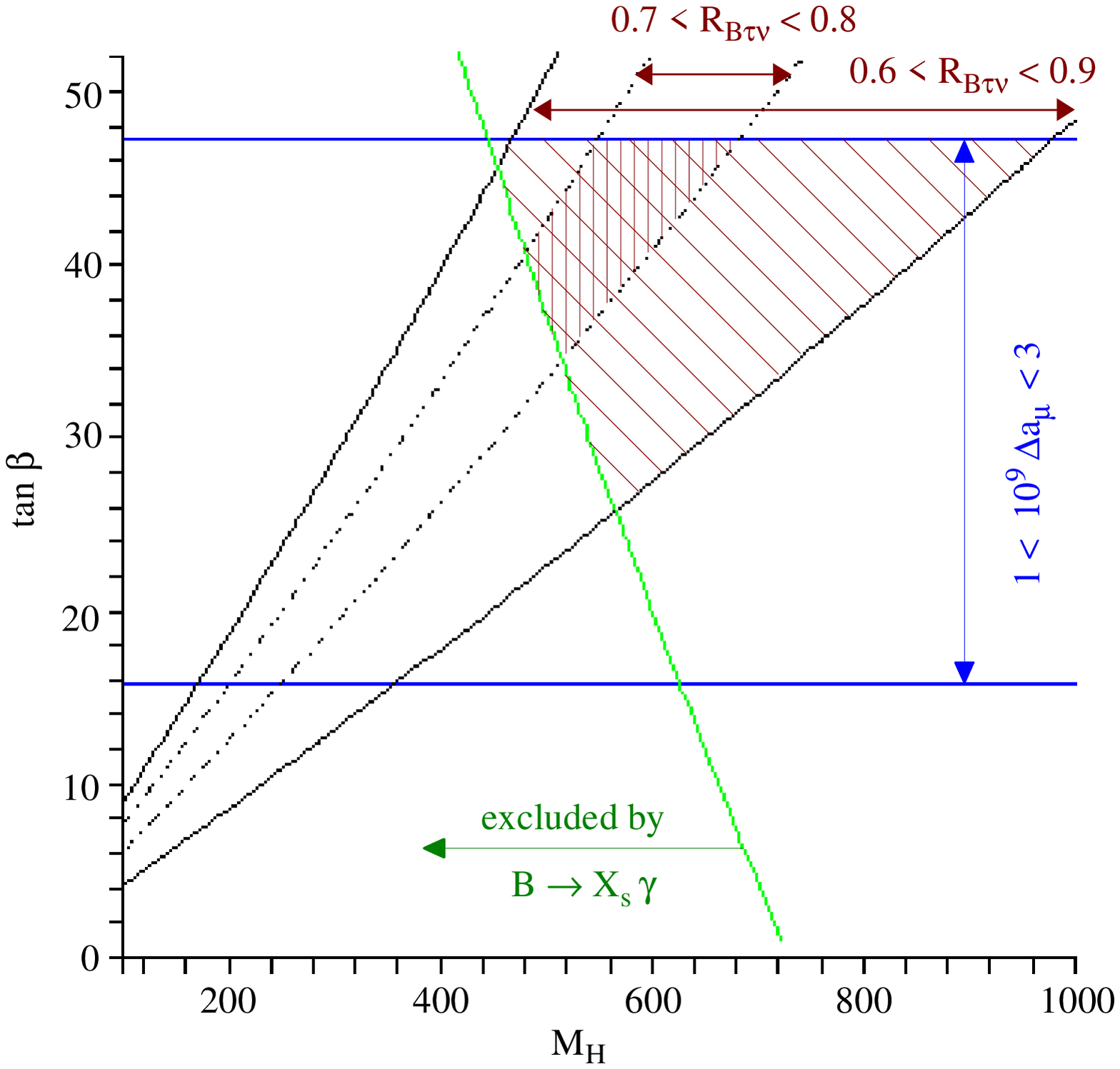}
\hspace{-0.3 cm}
\includegraphics[scale=0.46,angle=-90]{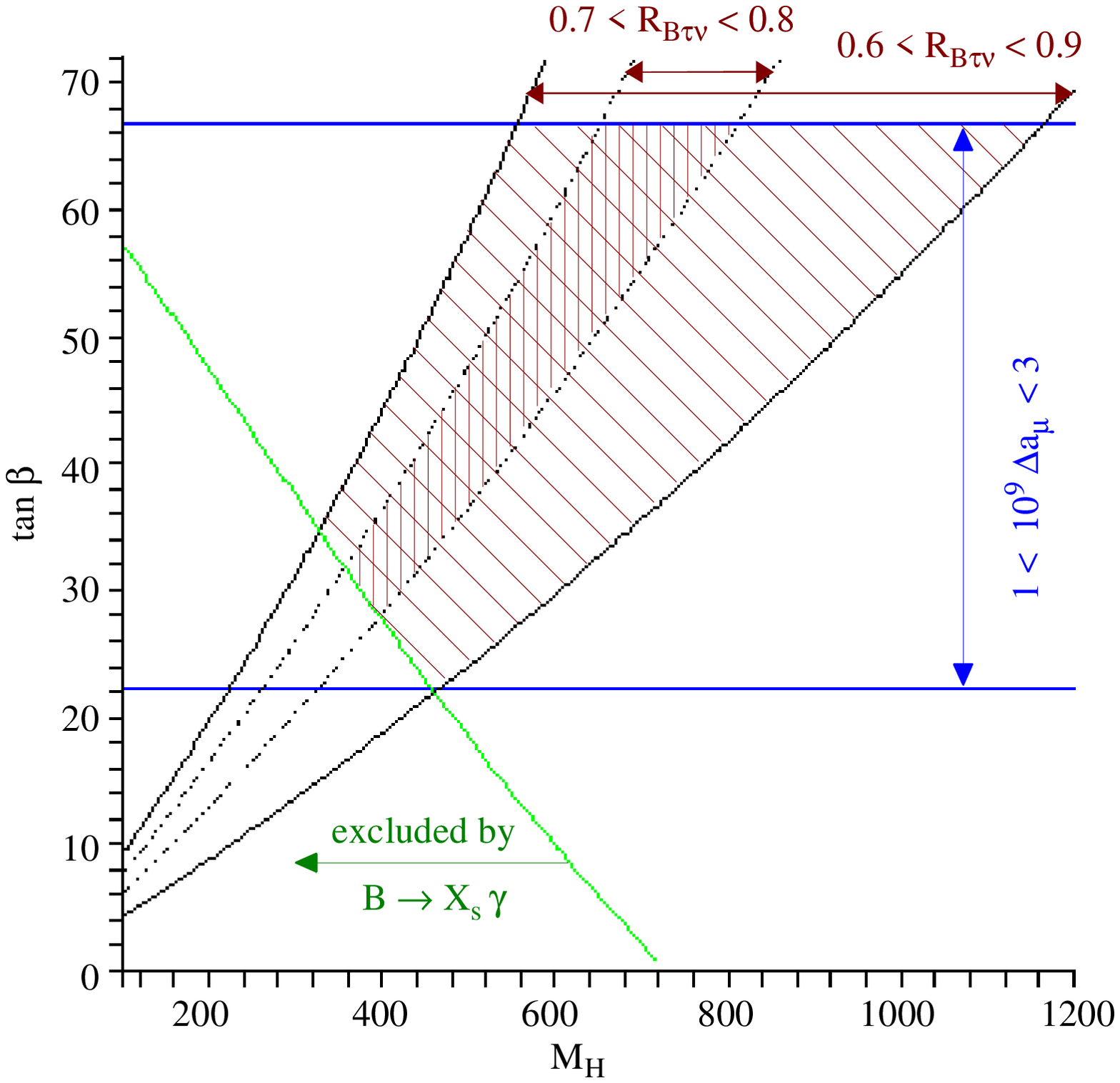}
\hspace{-0.3 cm} \\
\vspace{1.0 cm}
\hspace{-0.3 cm}
\includegraphics[scale=0.46,angle=-90]{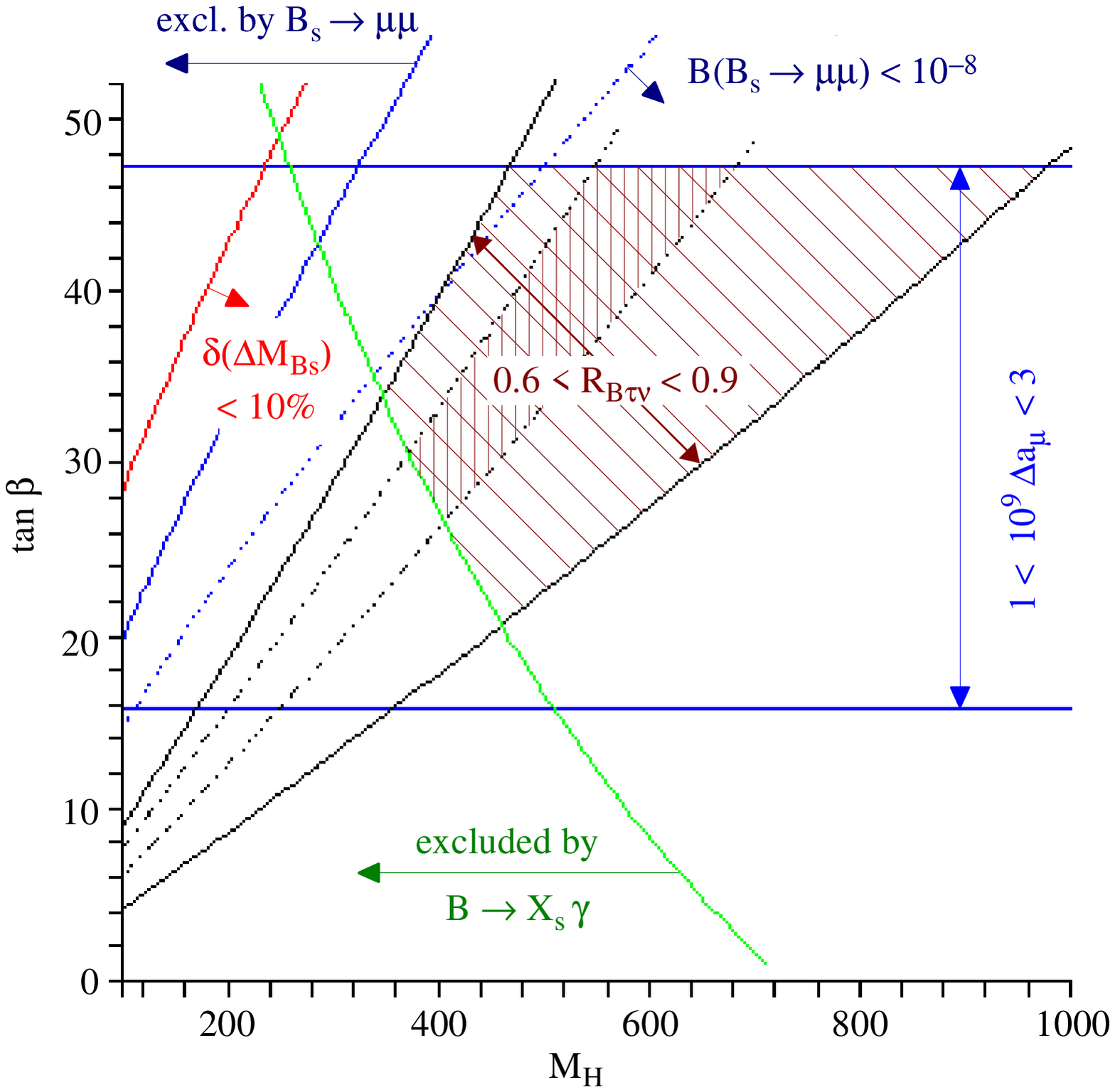}
\hspace{-0.3 cm}
\includegraphics[scale=0.46,angle=-90]{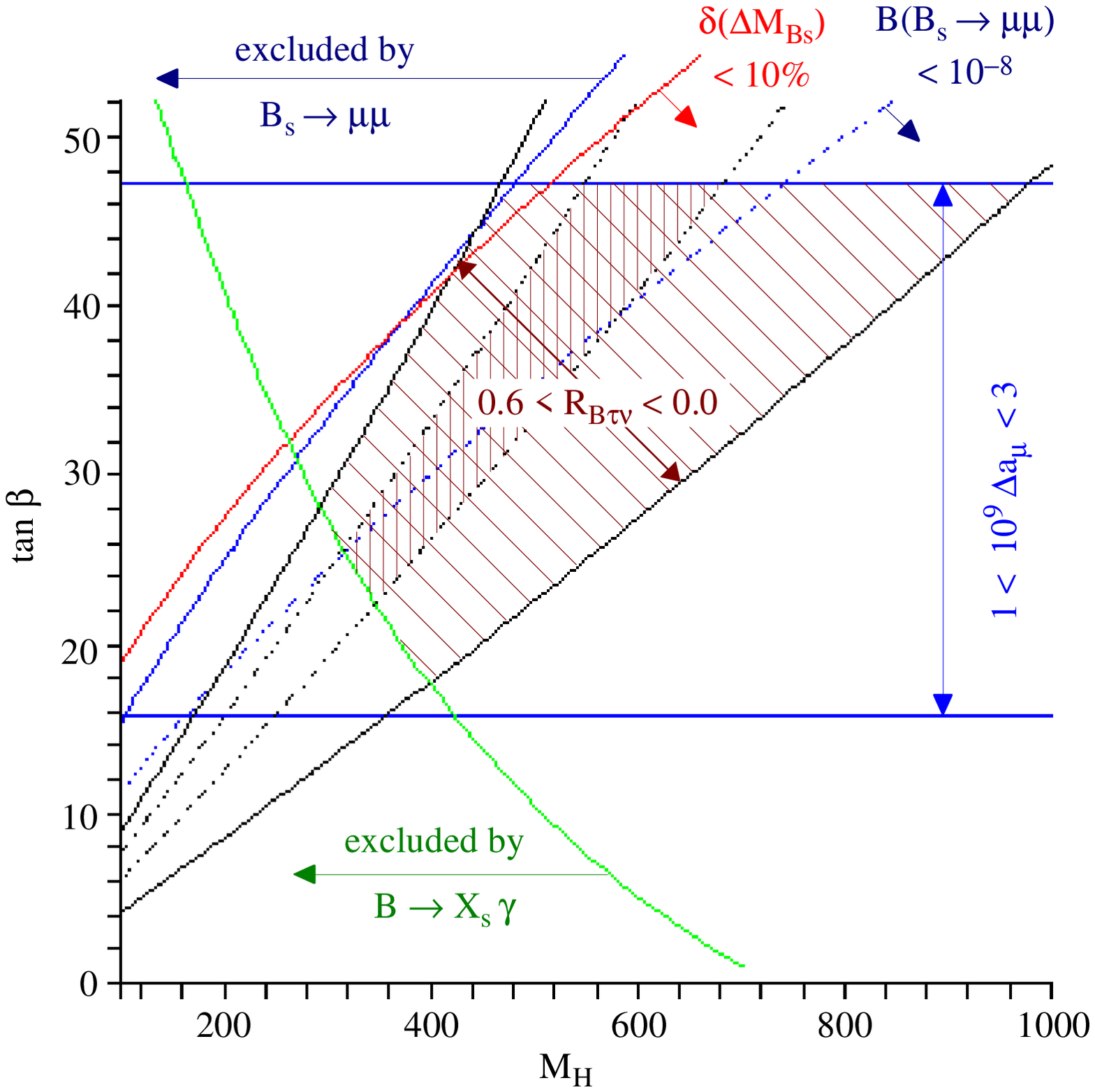}
\hspace{-0.3 cm} \\
\vspace{1.0 cm}
\caption{\label{fig:mu500} 
$B$-physics observables and $(g-2)_\mu$ in the $M_{H^{\pm}}$--$\tan\beta$ plane. 
The four plots correspond to:  $[\mu,A_U]=[0.5,0]$~TeV (upper left);
$[\mu,A_U]=[1,0]$~TeV (upper right); $[\mu,A_U]=[0.5,-1.0]$~TeV (lower left); 
$[\mu,A_U]=[0.5,-2.0]$~TeV (lower right). The exclusion regions for 
$\BR(B_s\to \mu^+\mu^-)$ and $\BR(B_s \to X_s\gamma)$ correspond to
the limits in Eqs.~(\ref{eq:Bll_lim}) and (\ref{eq:bsg_lim}), 
respectively (see main text for more details).}
\end{center}
\end{figure}

\begin{figure}[p]
\begin{center}
\hspace{-0.3 cm}
\includegraphics[scale=0.46,angle=-90]{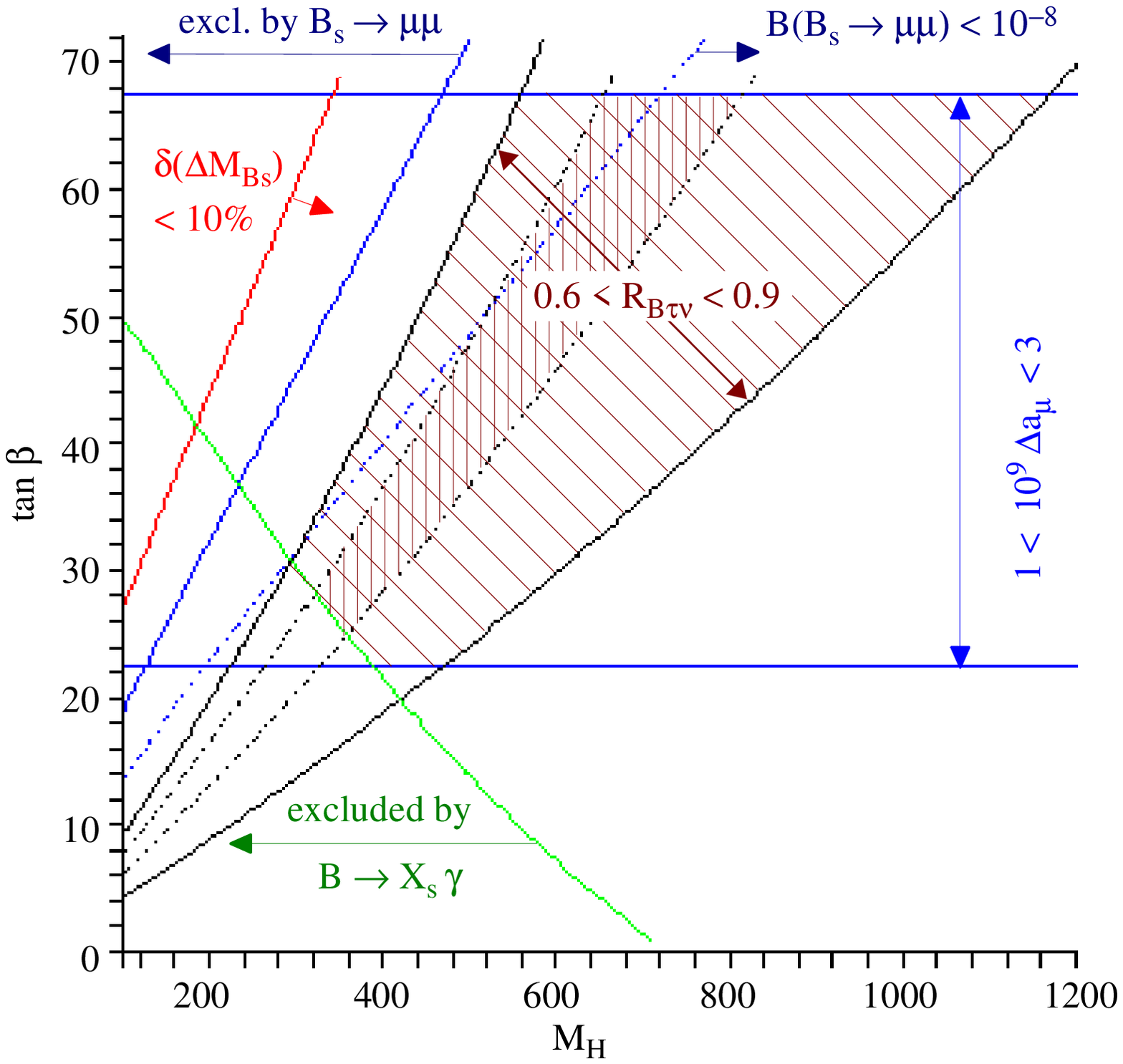}
\hspace{-0.3 cm}
\includegraphics[scale=0.46,angle=-90]{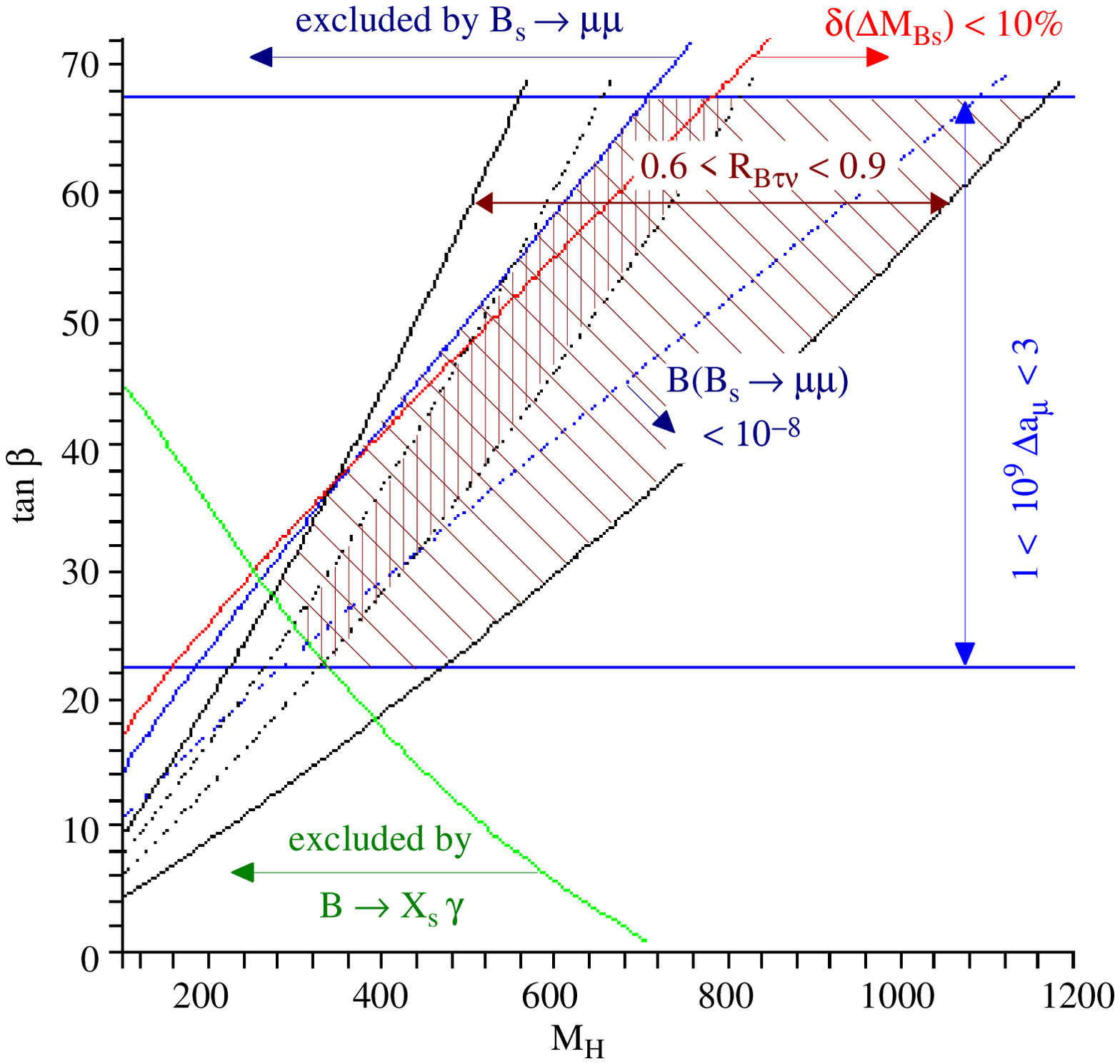}
\hspace{-0.3 cm} \\
\vspace{1.0 cm}
\hspace{-0.3 cm}
\includegraphics[scale=0.46,angle=-90]{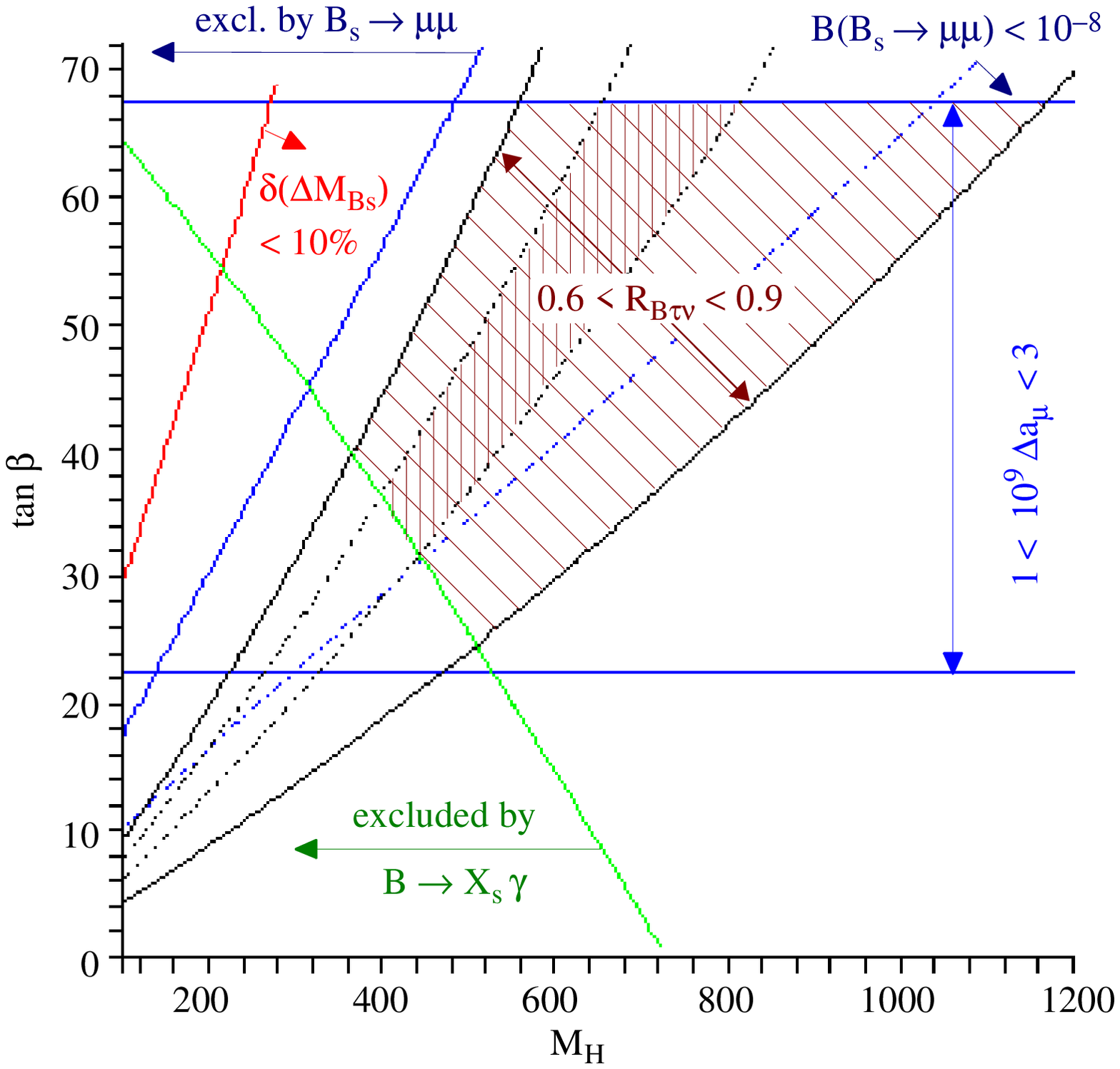}
\hspace{-0.3 cm}
\includegraphics[scale=0.46,angle=-90]{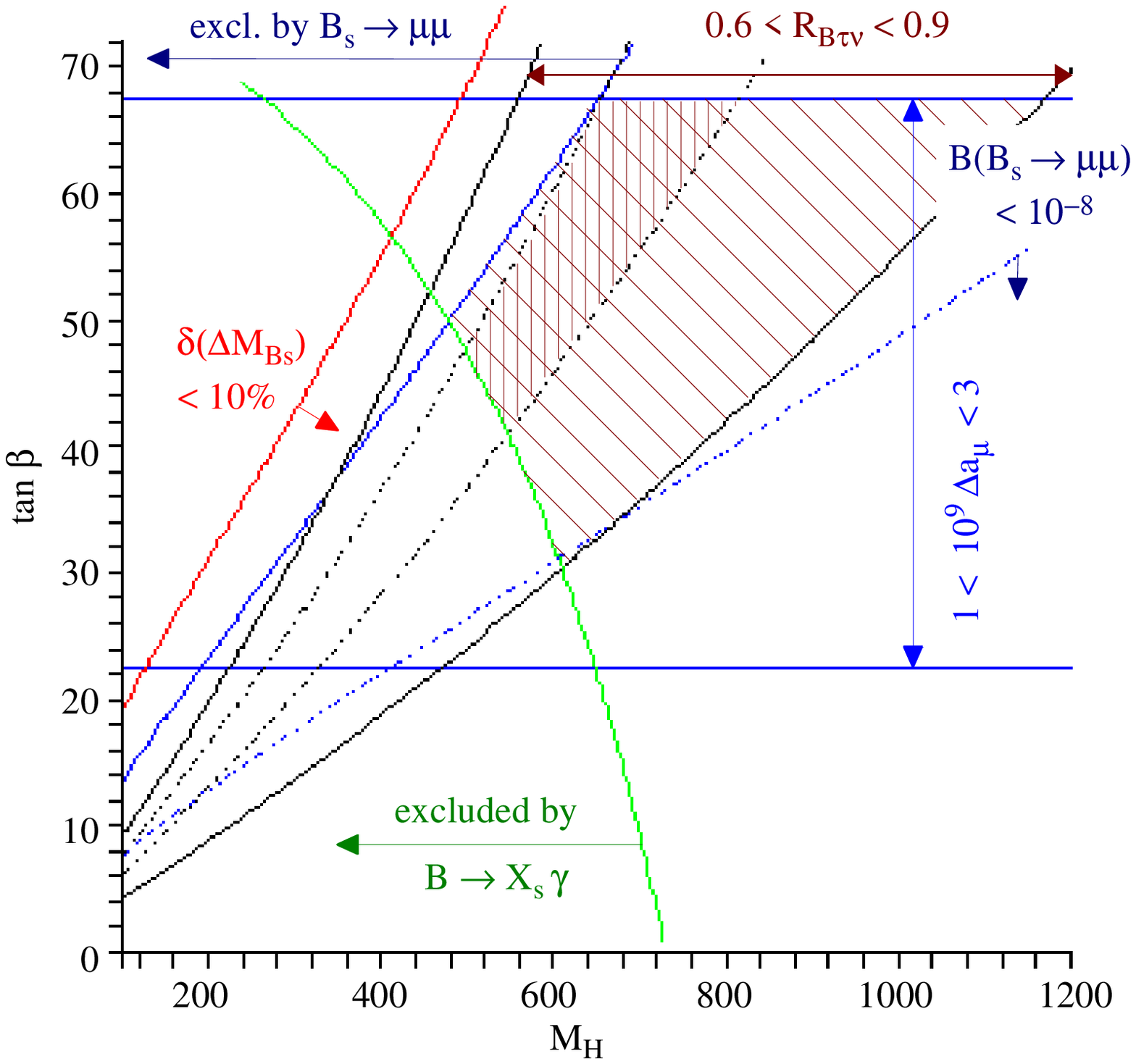}
\hspace{-0.3 cm} \\
\vspace{1.0 cm}
\caption{\label{fig:mu1000}
$B$-physics observables and $(g-2)_\mu$ in the $M_{H^{\pm}}$--$\tan\beta$ plane. 
The four plots correspond to:  $[\mu,A_U]=[1.0,-1.0]$~TeV (upper left);
$[\mu,A_U]=[1.0,-2.0]$~TeV (upper right); $[\mu,A_U]=[1.0,1.0]$~TeV (lower left); 
$[\mu,A_U]=[1.0,2.0]$~TeV (lower right). The exclusion regions for 
$\BR(B_s\to \mu^+\mu^-)$ and $\BR(B_s \to X_s\gamma)$ correspond to
the limits in Eqs.~(\ref{eq:Bll_lim}) and (\ref{eq:bsg_lim}), 
respectively (see main text for more details).}
\end{center}
\end{figure}

\section{Discussion}
\label{sect:discuss}
 
The correlations among the various observables discussed 
in the previous sections are illustrated in 
Figures~\ref{fig:mu500} and \ref{fig:mu1000}. Here we show 
the regions in the $M_{H^{\pm}}$--$\tan\beta$ plane which 
could give rise to a detectable deviation from the SM 
in  $\BR(\Btaun)$ and $(g-2)_\mu$, while satisfying the 
present experimental constraints from 
$\BR(B\to X_s \gamma)$, $\Delta M_{B_s}$ and 
$\BR(B_{s}\to \mu^+\mu^-)$. 

All plots have been obtained setting $\msq = 1$ TeV,
$\msl = 0.5$ TeV,  $M_2=0.3$ TeV, $M_1=0.2$ TeV, 
and changing the values of the two key parameters, $\mu$ (0.5  or 1 TeV) 
and $A$ (0, $\pm1$, $\pm2$ TeV), as indicated in the captions. 
We have explicitly checked that the structure of the plots remains 
essentially unchanged for variations of $M_2$ and $M_1$ in the 
range 0.2--0.5 TeV. Thus these plots can be considered 
as representatives of a wide area of the parameter space
(for squark masses in the TeV range). Note that, with the 
exception of $(g-2)_\mu$ (which depends on $\msl$) 
and $\BR(B\to X_s \gamma)$ (which depends on $\msq$), 
the other observables are completely 
independent from the absolute sfermion mass scale.

The dashed areas denote regions which yield 
a suppression of $\BR(\Btaun)$ of (10--40)\% or 
(20--30)\% (inner region), with a contribution to 
$\Delta a_{\mu}$ in the $1\sigma$ range defined 
by Eq.~(\ref{eq:amu_exp}). The exclusion regions 
from $\BR(B_{s}\to \mu^+\mu^-)$ and 
$\BR(B\to X_s \gamma)$ correspond to the 
bounds in  Eqs.~(\ref{eq:Bll_lim}) and (\ref{eq:bsg_lim}), 
respectively.
In all plots we have also indicated which is 
the impact of a $10\%$ bound on $\Delta M_{B_s}$ 
(i.e.~no more than 10\% suppression with respect 
to the SM), and the possible future impact 
of a more stringent bound on $\BR(B_{s}\to \mu^+\mu^-)$.

A list of comments follows:
\begin{itemize}

\item 
In all cases there is a wide allowed region of 
the parameter space with sizable (measurable) effects 
in $\BR(\Btaun)$, which would also provide a natural 
explanation of the $(g-2)_{\mu}$ problem. The only scenario 
where this does not happen is for $A_U \gsim 0.5$ TeV 
and $\mu \lsim 0.5$ TeV (plot not explicitly shown), 
where the $\BR(B\to X_s \gamma)$ constraint becomes 
particularly stringent.\footnote{~Note that in all cases
we have set a severe limitation on possible non-standard 
contributions to $\BR(B\to X_s \gamma)$: the allowed regions 
are determined by the upper bound in Eq.~(\ref{eq:bsg_lim}),
i.e.~by an increase of $\BR(B\to X_s \gamma)$ not exceeding $15\%$.}
\item 
If we require that charged Higgs effects account for 
a suppression of $\BR(\Btaun)$ of at least $10\%$, then we naturally 
have large SUSY effects in $(g-2)_{\mu}$ (except for unnaturally 
heavy sleptons). The viceversa is also true, but only 
if $M_{H^\pm}$ is sufficiently light ($M_{H^\pm} \lsim 600$ GeV),
or if we lower the ``detectability threshold'' of non-standard 
effects in  $\BR(\Btaun)$ to a few \%.

\item 
The $A_U=0$ case (Figure~\ref{fig:mu500}, upper plots) 
is shown only for illustrative purposes, to demonstrate that 
if $A_U$ is sufficiently small there is no connection 
between  $\BR(\Btaun)$, 
$\BR(B_{s}\to \mu^+\mu^-)$ and $\Delta M_{B_s}$. 
We do not consider this scenario very appealing because 
of the too light $m_{h^0}$ (see Section \ref{sect:mhh}).

\item
In the interesting cases with large $A_U$, the present 
data on $\BR(B_{s}\to \mu^+\mu^-)$ and $\Delta M_{B_s}$
imply only marginal limitations of the selected 
regions for $\BR(\Btaun)$ and  $(g-2)_{\mu}$. 
However, in all cases but for $[\mu,A_U]=[0.5,1]$~TeV
(Figure~\ref{fig:mu500}, lower left plot), 
a future limit on $\BR(B_{s}\to \mu^+\mu^-)$ 
at the  $10^{-8}$ level would cut a large fraction 
of the interesting region. In particular, 
we can conclude that if charged Higgs effects 
account for a suppression of $\BR(\Btaun)$ 
of at least $20\%$, then it is very likely 
that $\BR(B_{s}\to \mu^+\mu^-)$ exceeds $10^{-8}$ 
(to be compared with the SM level of 
$\approx 3.5 \times 10^{-9}$).

\item
In the scenario of maximal mixing ($A_U/\msq \approx 2$) 
and negative $A_U$, the $\Delta M_{B_s}$ bound is always 
at the border of the dashed areas. This implies that in this
scenario, which is quite interesting given the heavy $m_{h^0}$
and the effective cancellation of non-standard effects 
in $B\to X_s \gamma$,  $\Delta M_{B_s}$ can receive 
a small but non negligible suppression ($\lsim 10\% $) 
compared to its SM expectation. A clear evidence of this 
effect, together with a clear evidence of a larger 
suppression (20--40\%) in $\BR(\Btaun)$, would represent the 
unambiguous signature of this scenario.
\end{itemize}

\subsection{Other signatures in $P\to\ell\nu$ decays}

Besides $\BR(\Btaun)$,  $\BR(B_{s,d}\to \ell^+\ell^-)$
and $(g-2)_\mu$, which represent the most promising probes
of the MSSM scenario with heavy squarks and  large $\tan\beta$, 
specific signatures of this framework can be obtained by 
means of a systematic analysis of all $P\to\ell\nu$ modes:
\begin{description}
\item[$\underline{R_{P\ell\nu}}$]
In absence of sizable sources of flavour violation in the lepton 
sector, the expression in Eq.~(\ref{eq:Btn}) holds for all the $B$ purely 
leptonic decays. The corresponding expressions for the $K \to \ell \nu$ 
channels are obtained with the replacement $m_B \to m_K$, while for 
the $D \to \ell \nu$ case $m^{2}_{B} \to (m_s/m_c) m^{2}_{D}$. 
It is then easy to check that a $30\%$ suppression of $\BR(B \to \tau \nu)$ 
should be accompanied by a  $0.3\%$ suppression 
(relative to the SM) in $\BR(D \to \ell \nu)$ (see Ref.~\cite{dtnu}) 
and $\BR(K \to \ell \nu)$.
At present, the theoretical uncertainty on the corresponding decay 
constants does not allow to observe such effects.
However,  given the excellent experimental resolution on  
$K \to \mu \nu$ \cite{Kl2KLOE} and the recent progress from the lattice 
on kaon semileptonic form factors~(see e.g.~Ref.~\cite{Hashimoto}),
the identification of tiny deviations from the SM in this channel 
(compared to the SM prediction inferred from $K_{e 3}$ modes
\footnote{~To be more specific, the charged-Higgs exchange implies a $(0.1-0.2)\%$ suppression of the value of $|V_{us}|$ extracted from 
$K_{\mu 2}$ with respect to the one determined from $K_{e 3}$ decays.
This possibility is perfectly compatible (even slightly favoured \ldots) 
with present data \cite{Blucher}.}) is not hopeless in a future perspective.
\item[$\underline{R_P^{\ell/ \tau}}$]
As pointed out in Ref.~\cite{kl2}, if the model contains sizable
sources of flavour violation in the lepton sector (possibility 
which is well motivated by the large mixing angles in the neutrino sector), 
we can expect observable deviations  from the SM  also in the ratios 
\be
R_P^{\ell_1/\ell_2} = \frac{ \BR(P\to \ell_1 \nu) }{ \BR(P \to \ell_2 \nu)}~.
\ee
The lepton-flavour violating (LFV) effects can be quite 
large in $e$ or $\mu$ modes, while in first approximation 
are negligible in the $\tau$ channels. In particular, the leading parametric 
dependence of the most interesting $\BR(B \to \ell \nu)$ ratios is described by 
the following universal expression 
\bea
   \left(R_B^{\ell/\tau}\right)^{\rm MSSM}_{\rm LFV} =
  \left(R_B^{\ell/\tau}\right)^{\rm SM}
\left[1+\frac{1}{R_{B\tau\nu}} \left(\frac{m^{4}_{B}}{M^{4}_{H^{\pm}}}\right)
\left(\frac{m^{2}_{\tau}}{m^{2}_{\ell}}\right)|\Delta^{\tau \ell}_{R}|^2
\frac{\tan^6\beta}{(1+\epsilon_0 \tan\beta)^2} \right]~,\quad  \label{eq:LFV}
\eea
where the one-loop effective couplings $\Delta^{\tau \ell}_{R}$ 
can reach $\cO(10^{-3})$ \cite{kl2}. In the most favorable scenarios,
taking into account the constraints from LFV $\tau$ decays \cite{Paride},
Eq.~(\ref{eq:LFV}) implies spectacular order-of-magnitude enhancements 
for $R_B^{e/\tau}$ and  $\cO(10\%)$ deviations 
from the SM in $R_B^{\mu/\tau}$ (a detailed discussion about 
these effects is beyond the scope of the present work and
will be presented in a forthcoming publication).
\end{description}

\newpage

\section{Conclusions}

The observation of the $\Btaun$ transition \cite{Btn_Belle}
represents a fundamental step forward towards a deeper
understanding of both flavour and electroweak dynamics. 
The precise measurement of its decay rate could provide 
a clear evidence of a non-standard Higgs sector 
with large $\tan\beta$ \cite{Hou}. In this work we have analysed 
the interesting correlations existing between
this observable and $\BR(B\to X_s \gamma)$, 
$\Delta M_{B_s}$, $\BR(B_{s,d}\to \ell^+\ell^-)$, and 
$(g-2)_\mu$, in the large $\tan\beta$ regime of the MSSM.
We have shown that this scenario is particularly interesting,
especially in the limit of heavy squarks and trilinear 
terms. In this framework one could naturally accommodate 
the present (non-standard) central values of both $\BR(\Btaun)$
and $(g-2)_\mu$, explain why the 
lightest Higgs boson has not been observed yet,
and why no signal of new physics has been
observed in $\BR(B\to X_s \gamma)$ 
and $\Delta M_{B_s}$.

One of the virtues of the large $\tan\beta$ regime of the 
MSSM, with MFV and heavy squarks, is its naturalness in 
flavor physics and in precise electroweak tests. 
As we have shown, 
no fine tuning is required to accommodate the precise SM-like 
results in $\BR(B\to X_s \gamma)$ and $\Delta M_{B_s}$. 
On the other hand, the scenario could clearly be distinguished by 
the SM with more precise results on  $\BR(\Btaun)$, 
and possibly on $\BR(B_{s,d}\to \ell^+\ell^-)$. In particular,
we have discussed how to decrease the theoretical/parametric
uncertainties in the SM prediction of $\BR(\Btaun)$
by normalizing this observable to $\Delta M_{B_d}$. 
Moreover, we have shown that in the most favorable scenarios 
for $m_{h^0}$ (i.e.~for $A_U > \msq$), 
a (20--30)\% supression of $\BR(\Btaun)$ is accompanied 
by enhancements of the $B_{s,d}\to \ell^+\ell^-$ rates 
by more than a factor of 3 compared to the 
corresponding SM expectations.

The observables $\BR(\Btaun)$,  $\BR(B_{s,d}\to \ell^+\ell^-)$
and $(g-2)_\mu$ can be considered as the most promising 
low-energy probes of the MSSM scenario with heavy squarks and large $\tan\beta$. 
Nonetheless, as discussed in Section~\ref{sect:discuss}, 
interesting consequences of this scenario could 
possibly be identified also in $\Delta M_{B_s}$ 
and in other $P\to \ell\nu$ modes. 
In particular, if  $A_U$ is large and negative 
we expect a $\approx 5 \%$ suppression of  $\Delta M_{B_s}$
with respect to the SM expectation (possibility which 
is certainly not excluded by present data \cite{DMs_CDF}). 
A non-standard effect of this magnitude in  $\Delta M_{B_s}$
is very difficult to be detected, but it is not hopeless in view 
of improved lattice data (see e.g.~Ref.~\cite{Hashimoto})
and more refined CKM fits (see e.g.~Ref.~\cite{Bona,Ligeti}).
The model also predicts a few per-mil suppression of 
the $|V_{us}|$ value extracted from $K_{\mu 2}$
(compared to the $K_{e 3}$ one). 
Finally, if the slepton sector contains sizable 
sources of flavour violation,
we could even hope to observe large violations 
of lepton universality in the ratios 
$\BR(B\to \mu \nu)/\BR(B \to \tau \nu)$ and 
$\BR(B\to  e \nu)/\BR(B \to \tau \nu)$,
as well as few per-mil effects in 
$\BR(K\to e \nu)/\BR(K \to \mu \nu)$ \cite{kl2}.

\end{document}